\documentclass[prd,showpacs,preprintnumbers,floatfix,superscriptaddress,10pt]{revtex4}
\usepackage[mathscr]{eucal}
\usepackage{color}
\usepackage[dvips]{graphicx}
\usepackage{epsf}
\usepackage{bm}
\usepackage{epsfig}
\usepackage{feynmf}
\usepackage{amsmath}

\newcommand{\be}{\begin{equation}}
\newcommand{\ee}{\end{equation}}
\newcommand{\ba}{\begin{eqnarray}}
\newcommand{\ea}{\end{eqnarray}}

\newcommand{\bea}{\begin{eqnarray}}
\newcommand{\eea}{\end{eqnarray}}

\newcommand{\f}{f_p^0}

\begin{document}

\title{Thermal conductivity  due to phonons in the core of  superfluid neutron stars}

\author{Cristina Manuel}
\email{cmanuel@ieec.uab.es}
\affiliation{Instituto de Ciencias del Espacio (IEEC/CSIC) Campus Universitat Aut\`onoma de Barcelona, Facultat de Ci\`encies, Torre C5, E-08193 Bellaterra (Barcelona), Catalonia, Spain}

\author{Sreemoyee Sarkar}
\email{sreemoyee.sarkar@tifr.res.in}
\affiliation{High Energy Nuclear and Particle Physics Division, Saha Institute of Nuclear Physics,
1/AF Bidhannagar, Kolkata-700 064, India}

\author{Laura Tolos}
\email{tolos@ice.csic.es}
\affiliation{Instituto de Ciencias del Espacio (IEEC/CSIC) Campus Universitat Aut\`onoma de Barcelona, Facultat de Ci\`encies, Torre C5, E-08193 Bellaterra (Barcelona), Catalonia, Spain}
\affiliation{Frankfurt Institute for Advances Studies. Johann Wolfgang Goethe University, Ruth-Moufang-Str. 1,
60438 Frankfurt am Main, Germany }


\begin{abstract}
{We compute the contribution of phonons to the thermal conductivity in the core of superfluid neutron stars.
 We use effective field theory techniques to extract the phonon scattering rates, written as a function of
the equation of state of the system. We also calculate the phonon dispersion law beyond linear order, which depends on the gap of superfluid neutron matter. With all these ingredients,  we solve the  Boltzmann equation numerically using a variational approach. We find that the thermal conductivity $\kappa$ is dominated by combined
small and large angle binary collisions. As in the color-flavor-locked superfluid, we find that 
 our result can be well approximated by  $\kappa \propto 1/ \Delta^6$  at low T, where $\Delta$ is the neutron gap, 
 the constant of proportionality depending on the  density. We further comment on the possible relevance of electron and superfluid
phonon collisions in obtaining the total contribution to the thermal conductivity in the core of superfluid neutron stars.
}
\end{abstract}

\date{\today}
\maketitle

\section{Introduction}
\label{Intro}

In this paper we compute the superfluid phonon contribution to the thermal conductivity ($\kappa$) in the core of superfluid neutron stars.
This is a follow up of other publications  where the superfluid phonon contribution to the shear viscosity \cite{Manuel:2011ed}
or bulk viscosity coefficients  \cite{Manuel:2013bwa} for this system  have been computed, with the aim of assessing its effect on the dynamics of the hydrodynamical modes
of the star \cite{Manuel:2012rd}.

The cooling of a neutron star depends on the rate of neutrino emission, the
photon emission rate, the specific heat and  also in young stars on the thermal conductivity \cite{Pethick:1991mk,Yakovlev:2004iq}. As all these quantities depend on the
microscopic processes occurring inside the star, it has been recognized that the measurement of the cooling rate of a neutron star could give some hints on its composition.
That is why  computing the thermal conductivity in a compact star is so relevant. Further, the thermal conductivity  is also needed to get the relaxation of any heat
flux produced by any dissipative process in the star.

So far the thermal conductivity in the core of neutrons stars  has been computed assuming
both normal and superfluid neutron matter phases \cite{Flowers,Baiko:2001cj,Shternin:2007ee}. In the last case, it has been claimed that the thermal conductivity is dominated by
electron collisions (and muons, for sufficiently large densities inside the core of the star). In this paper we compute the contribution to the
thermal conductivity of the superfluid phonon, which is a collective mode which appears due to the onset of superfluidity. We leave for a future project a careful study of
the interactions of the superfluid phonons with the electrons, and their impact on the transport properties of the core of neutron stars.

We use effective field theory techniques to assess the relevant superfluid phonon scattering amplitudes. It is known that at leading order (LO) in an energy and momentum expansion
the self-phonon interactions can be determined with the knowledge of the equation of state (EoS) of the superfluid \cite{Son:2005rv}. We use this fact to determine the collision term that enters
the Boltzmann equation, which we solve in order to determine 
 the value of the thermal conductivity $\kappa$. Technically, the problem is totally analogous to the computation of
the superfluid phonon contribution to the thermal conductivity in color-flavor locked quark matter \cite{Braby:2009dw}, or for the cold Fermi atoms in the unitarity limit \cite{Braby:2010ec}, the main difference arises from the fact that, since the systems are different, the value of the phonon couplings and dispersion laws differ.

 In this work, as in our previous studies, we consider a simplified model
of neutron star made up by neutrons, protons and electrons, using a causal parametrization of the Akmal, Pandharipande and Ravenhall (APR for short) EoS \cite{ak-pan-rav,Heiselberg:1999mq}  to describe $\beta$-stable nuclear matter inside the star, and get from it all the phonon self-couplings.   For the superfluid gap, we use two different parametrizations in order to consider the cases of strong and weak superfluidity \cite{Andersson:2004aa}. We also note, as previously found out for other superfluid systems \cite{Khala,Braby:2009dw}, that if we take into account the phonon dispersion law at LO
the thermal conductivity strictly vanishes. Knowledge of the phonon dispersion law beyond LO is then needed. Thus, we first compute such a correction to the linear dispersion law
by considering s-wave neutron pairing, and assume that the phonon dispersion law remains the same in the p-wave phase that is supposed to prevail in most part of the superfluid core.

At this stage one should mention that superfluid phonons may be overdamped or even disappear
for some temperatures $T$ \cite{Leggett:1966zz, Leinson:2010ru}. Leggett determined the overdamping of these collective modes in superfluid fermionic systems by analyzing the associated Landau parameters within the Fermi liquid theory. However, for the present case of the APR EoS, we cannot explore this possibility since we cannot extract the corresponding Landau parameters as the nucleon-nucleon effective interaction is not available for the given EoS model. Thus, we will not take into account the possible disappearance of superfluid phonons leaving for a future project an  alternative independent study on the overdamping of superfluid phonons in the interior of neutron stars.


This paper is structured as follows. In Sec.~\ref{SPsections} we discuss the relevant ingredients of the phonon dynamics for our computation. In Sec~\ref{dispersion-sec} we
 obtain the superfluid phonon dispersion law beyond  linear order, and in Sec.~\ref{phononcollisions} we
 show the relevant phonon collisions which contribute to the thermal conductivity. 
In Sec.~\ref{sec:variational} we explain the variational approach we use to solve the Boltzmann equation, which is the same as in Ref.\cite{Braby:2009dw}. In Sec.~\ref{ScalingSec} we present some  arguments to derive the scaling behavior  of the thermal conductivity with the temperature and
gap, respectively, and in Sec.~\ref{numericalsec} we show  our numerical results. In  Sec.~\ref{conclusions} we present a
comparison between the electron-muon and phonon contributions to the
thermal conductivity  and discuss that a careful study of the electron-phonon interactions is still needed to fully determine the value of $\kappa$. We use natural units ($\hbar = c= k_B=1$) throughout the manuscript, although we present our results in c.g.s. units.

\section{Superfluid phonons in the core of neutron stars}
\label{SPsections}

The superfluid phonon is a Goldstone mode which appears due to the spontaneous symmetry breaking of the $U(1)$ symmetry
induced by the appearance of a difermion condensate. The dynamics of the phonon can be completely determined from the microscopic 
physics, if one integrates out the heavy degrees of freedom of the system. The basic phonon interaction rates can be derived using effective field theory techniques. The effective phonon Lagrangian is  presented as an expansion in derivatives (or momenta and frequencies) of the phonon field, the terms of this expansion being restricted by symmetry considerations. The coefficients of the Lagrangian can be  computed from the microscopic theory, through a standard
matching procedure, and thus  they depend on the short range physics of the system under
consideration. The effective field theory assumes that the phonon momentum $k$ and  frequency $k_0$ are such that $k v_F,  k_0  \ll \Delta$, where $v_F$ is the Fermi velocity.
The expansion parameter of the theory is not a coupling constant, bur rather the values of $k v_F/ \Delta$ and $k_0 / \Delta$.

At LO in the derivative expansion, all the phonon self-couplings can be parametrized in terms of the speed of sound, the density of the superfluid, and derivatives of the speed of sound with respect to the density 
at $T=0$ \cite{Escobedo:2010uv}. Thus, the phonon physics at leading order is completely determined by the EoS of the superfluid. In this work, as in our previous studies, we consider 
a model of neutron star made up by neutrons, protons and electrons, using a causal parametrization of the APR EoS \cite{ak-pan-rav,Heiselberg:1999mq}. This EoS is widely used as benchmark for $\beta$-stable neutron star matter and,  from this EoS, we obtain all the phonon self-couplings. We refer the reader to Ref.~\cite{Manuel:2011ed} for more explicit details of the phonon Lagrangian and EoS used in this work.

For the computation of the superfluid phonon contribution to the thermal conductivity, the phonon dispersion law at next to leading order (NLO) is however required. This is so, as the thermal conductivity strictly vanishes if one uses
a linear dispersion law. 
In this Section  we compute the first correction beyond linear order to the phonon dispersion law, and later on we show the phonon collisions that we consider in the computation of the thermal conductivity.

\subsection{ Superfluid phonon dispersion law beyond linear order}
\label{dispersion-sec}

In this Subsection we will compute the phonon dispersion law, assuming s-wave pairing of the neutrons.
To study the main properties of the superfluid system, it is enough to consider a Lagrangian keeping only
 the interaction that is responsible for the formation of the gap, after  integrating out all the modes
 away from Fermi surface \cite{Shankar:1993pf,Polchinski:1992ed}. The relevant Lagrangian then reads
\begin{eqnarray}
{\cal L}_N &= & 
 \psi_n^\dagger ( i \partial_t - \frac{\nabla^2}{2 m^*} + \mu) \psi_n  - \frac{G}{2} \left( \psi_n^\dagger(x) \psi_n(x)\right) \left( \psi_n^\dagger(x) \psi_n(x)\right) + \cdots
\end{eqnarray}
where $\psi_n$ is the neutron wave-function and $\mu$ is the chemical potential associated to the neutrons.
Here $m^*$ and $G$ should be matched with the complete microscopic theory to correctly reproduce the value of
the Fermi momentum and gap, respectively.
 One can write 
$\psi_n = \psi_{n,0} e^{i \varphi}$,
where the phase of the neutron wave-function is the Goldstone field.
The kinetic term associated to $\psi_{n,0}$ is the same as that for the whole neutron wave-function, where one only
has to substitute the standard derivatives by covariant derivatives
\begin{equation}
 \psi_n^\dagger ( i \partial_t - \frac{\nabla^2}{2 m^*} + \mu) \psi_n =
 \psi_{n,0}^\dagger ( i (\partial_t + i {\tilde A}_0) - \frac{(\nabla + i {\bf \tilde A})^2}{2 m^*} + \mu) \psi_{n,0} \ ,
 \end{equation}
where ${\tilde A}_\mu = (\partial_t \varphi, \nabla \varphi)$ is written in terms of the Goldstone field $\varphi$.
Integrating out the field  $\psi_{n,0}$ one is left with an effective field theory  for the superfluid phonon, from which its dispersion law
can be obtained.

For this computation  we use the Nambu-Gorkov formalism (for a review see, e.g., Ref.~\cite{Schafer:2006yf}).
In the Nambu-Gorkov basis 
\bea
\label{ng_basis}
 \chi= \frac{1}{\sqrt 2}
 \left(\begin{array}{cc}
     \psi   \\
     \psi_c
     \end{array}\right), 
\eea
where $\psi_c = C\psi^{\star}$ and $C = i \sigma_2$ where $\sigma_2$ is a Pauli matrix,   the fermion propagator then becomes
\bea
 S(p_0, {\bf p} )=\frac{1}{p_0^2-\xi_p^2-\Delta^2}
 \left(\begin{array}{cc}
     p_0+\xi_p  & \Delta \\
     \Delta^{\star} & p_0-\xi_p
 \end{array}\right)  \ ,
 \label{ferm_ng_prop}
 \eea
where $\xi_p =\frac{p^2}{2m^*} - \mu$, and $\Delta$ is the gap parameter.

\begin{figure}
\begin{center}
\psfig{file=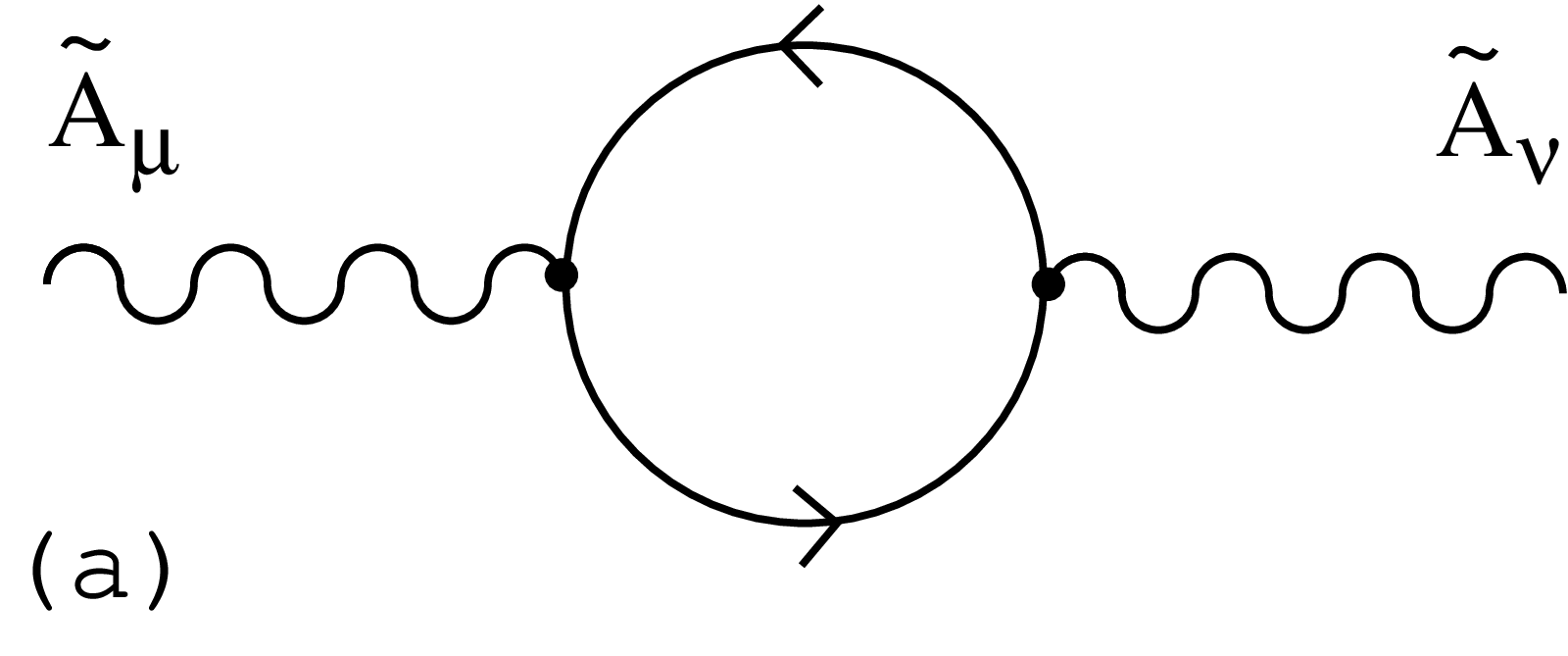,width=.40\hsize}
\hskip.05\hsize
\psfig{file=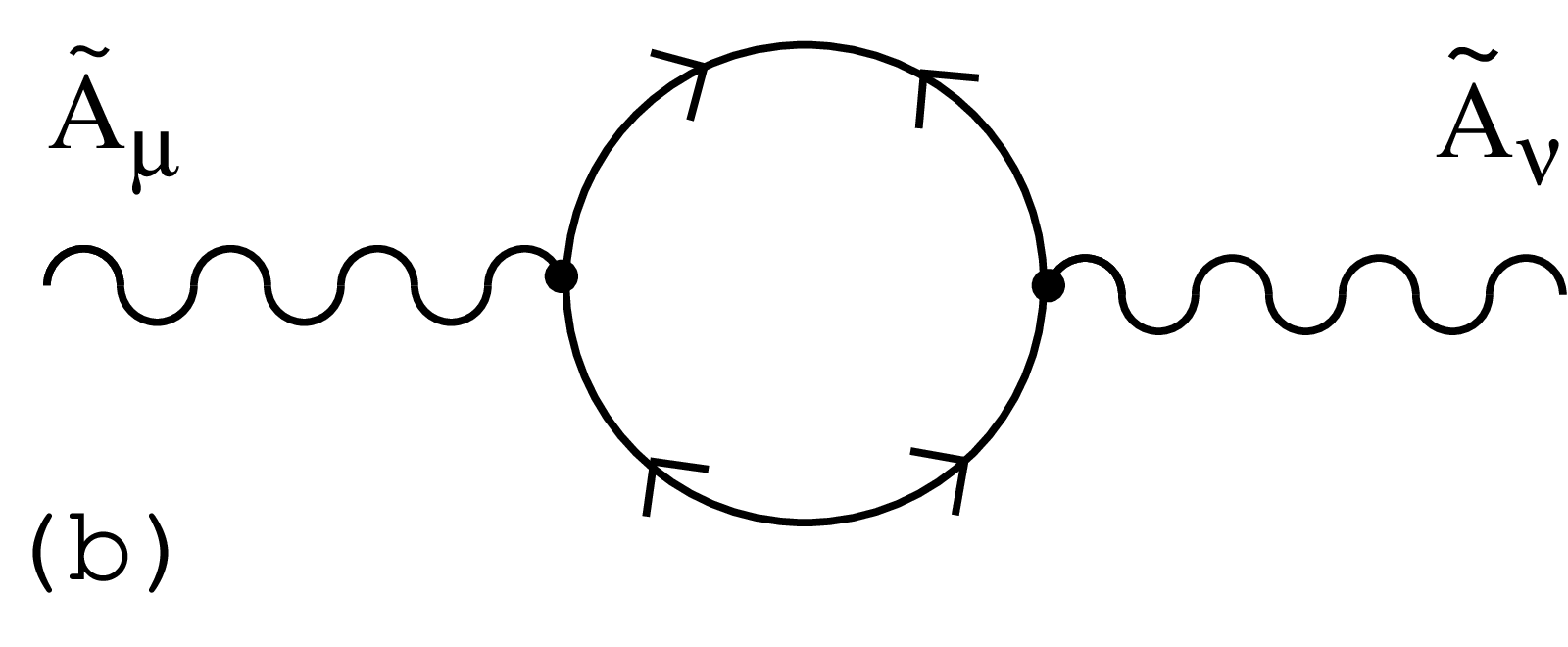,width=.40\hsize}
  \caption{ One-loop Feynman diagrams needed for the computation of the phonon dispersion law. In a) and b) the diagrams correspond to the
  diagonal/off-diagonal terms in the Nambu-Gorkov basis, respectively.}
\label{Polarization}
\end{center}
\end{figure}

The phonon dispersion law is then obtained by studying the solutions to the equation
$k^\mu k^\nu  \Pi_{\mu \nu}(k_0, {\bf k}) = 0$, where  $\Pi_{\mu \nu}$ is obtained after computing (see the one-loop Feynman diagrams displayed
in Fig.~\ref{Polarization})
\bea
\label{pimunu}
\Pi_{\mu \nu}(k_0, {\bf k}) =-i 
  \int\!\frac{d^4p}{(2\pi)^4} 
   \mbox{Tr}_{NG}\left[\Gamma_\mu S(p_0, {\bf p} )\Gamma_\nu S(p_0,-k_0 ,{\bf p - k})\right],
\eea  
where  $\mbox{Tr}_{NG}$ is a 
trace performed on the Nambu-Gorkov indices, and 
 $\Gamma_\mu = (\Gamma_0, \Gamma_i)$ is the vertex factor for the coupling of 
  the ${\tilde A}_\mu$  field in the Nambu-Gorkov basis, and can be expressed as
\bea
\label{ng_vert_00}
 \Gamma_0 = 
 \left(\begin{array}{cc}
     1  & 0 \\
     0 & -1
 \end{array}\right) \ , \qquad \Gamma_i = \frac{1}{2m^*}
 \left(\begin{array}{cc}
     p_i  & 0 \\
     0 & p_i
 \end{array}\right).
\eea

The computation of the one-loop diagram can be carried out analytically if one performs an expansion in the parameters 
$k_0/\Delta$ and $k v_F/\Delta$, see the Appendix \ref{App:Polarization} for more explicit details.
Then one finds that the phonon dispersion law can be expressed as  
\bea
\label{NLOdisp}
 E_k = c_s k\left(1+\gamma k^2\right) + \mathcal{O} (k^5)  \ ,
 \eea
 where 
 \bea
 c_s = \frac{v_F}{\sqrt{3}} \ , \qquad \gamma = - \frac{v_F^2}{45 \Delta^2} \ ,
 \eea
and $c_s$ is the speed of sound of the system.
 

In most part of the core of the neutron stars, neutrons are supposed to condense in a $^3P_2$ channel.
A similar computation to the one performed here could be used to determine the phonon dispersion law.
Then one should keep in the neutron Lagrangian the interaction responsible for the formation of the gap, using
the neutron propagator in the p-wave phase.  Such a computation has been carried out  in Ref.~\cite{Bedaque:2003wj}, but only the linear part of the phonon dispersion law has been
obtained. In this paper we will not compute the correction to the linear term  but simply assume that the
correction is expressed as  in the s-wave phase, using for $\Delta$ the angle-averaged value of the p-wave gap.

\subsection{Relevant phonon collisions for the thermal conductivity}
\label{phononcollisions}

\begin{figure}
\begin{center}
\hbox{\psfig{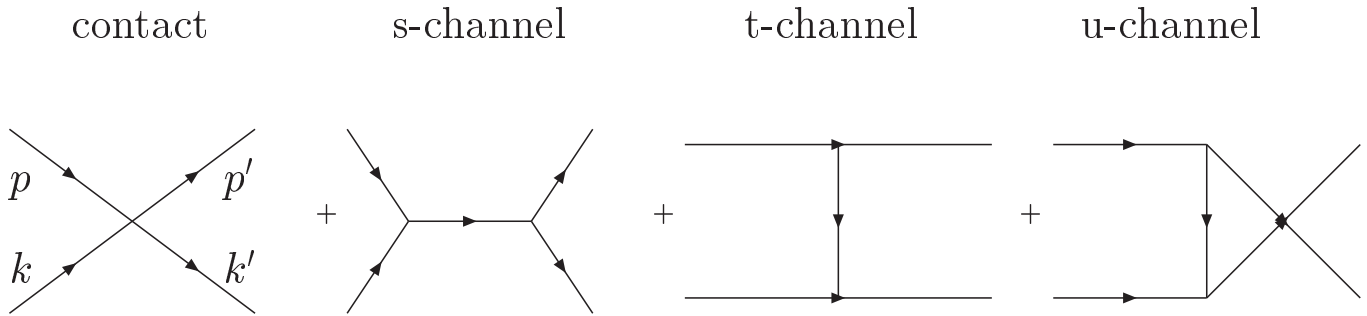}} \caption{
Feynman diagrams contributing to $2\leftrightarrow 2$ phonon
scattering processes.}
 \label{feyndiags}
\end{center}
\end{figure}

In this Subsection we briefly recall  the relevant phonon collisions which are needed for the computation of the thermal conductivity.
As the phonon dispersion law curves downward beyond linear order collisional processes of one phonon decaying into two or more phonons
are kinematically forbidden (see the Appendix A of Ref.~\cite{Manuel:2013bwa} for a brief discussion on this point). Thus, the first leading 
processes relevant for the thermal conductivity are binary collisions of phonons, see Fig.~\ref{feyndiags}.
Expressions for the different scattering amplitudes relevant for the binary collisions were explicitly given in Ref.~\cite{Manuel:2011ed}.
We will not write them here again, but comment on the change in the amplitudes we will be using here with respect to those used
in the computation of the shear viscosity of Ref.~\cite{Manuel:2011ed}.

It turns out that if one naively uses the scattering amplitudes obtained  from the LO effective field theory, then collinear
singularities arise in the evaluation of the different transport coefficients \cite{Manuel:2011ed,Manuel:2013bwa}.
 One then has to regulate the phonon propagators appearing in the different scattering amplitudes. For the computation of the shear viscosity,
we regulated the phonon propagator with the  one-loop thermal phonon damping \cite{Manuel:2011ed,Manuel:2004iv}. But we also checked that a different possible
regularization, using the non-linear phonon dispersion law, was possible leading to the same results. This is due to the fact that the shear viscosity is dominated by the large angle collisions, as the viscosity 
describes transport in the direction orthogonal to the flow.  In the computation of the thermal conductivity we see that the final result depends on large angle 
as well as on collinear (or small angle) collisions. 
 In the collinear regime, the phonon propagators in the scattering amplitudes becomes almost on-shell.  It has been checked that the imaginary part of the phonon self-energy vanishes when evaluated on-shell for $\gamma <0$
 \cite{Escobedo:2010uv}. Because in superfluid neutron matter we have found $\gamma < 0$, we are forced to regulate the phonon propagators appearing in the evaluation of the scattering amplitudes by taking the 
 phonon dispersion beyond linear order. That is, we will be using the same scattering amplitudes that were used in Ref.~\cite{Manuel:2011ed}, but with the phonon propagator expressed as
 \be
 G(k_0,{\bf k}) = \frac{1}{k_0^2 - E^2_k} \ , 
 \ee
with $E_k$ given by Eq.~(\ref{NLOdisp}).

 \section{Thermal conductivity}
 \label{thermal}
 
In this Section we present the results for the phonon contribution to the thermal conductivity in  superfluid $\beta$-stable neutron matter. We start by summarizing the variational formalism we use to solve the Boltzmann equation \cite{Braby:2009dw}.  We then present a scaling argument to obtain  the temperature and gap dependences of the thermal conductivity, which are determined by the leading contribution to the thermal conductivity coming from large and/or small angle phonon-phonon collisions. We finish this Section by showing the results of the numerical computations
of the thermal conductivity.
  
 \subsection{Variational computation of the thermal conductivity}
 \label{sec:variational}
 
 In hydrodynamics the thermal conductivity   $\kappa$ is the coefficient
 that relates the heat flux with a temperature gradient as
 \bea
{\bf  q}=-\kappa \nabla T \ .
\label{defi}
 \eea
 In kinetic theory  the heat flow can be expressed as
 \bea
 {\bf q}=\int \frac{d^3p}{(2\pi)^3} {\bf v}_p\, E_p \,\delta f_p \ ,
\label{ht_flux}
 \eea
 where 
 ${\bf v}_p=\partial E_P/ \partial{\bf  p}$ is the particle velocity, and $\delta f_p=f_p-f_p^0$ is the deviation of the distribution function 
 from the local thermal equilibrium given by
 \bea
 f_p^0=\frac{1}{e^{p_{\mu}u^{\mu}/T}-1}  \ ,
 \eea
and $u_{\mu}$ is the local fluid velocity.

 For the computation of the thermal conductivity one typically expresses the deviation from
equilibrium as
 \bea
\delta f_p = -\frac{f_p^0 (1+f_p^0)}{T^3} g(p)  {\bm p} \cdot {\bm\nabla} T\ ,
\label{df}
\eea
and $g(p)$ is a dimensionless function. Thus, the thermal conductivity can be expressed as
\bea
\kappa = \frac{1}{3 T^3} \int \frac{d^3 p}{(2\pi)^3}\, \f (1+\f) g(p)\, 
v_p\,E_p\, p.
\label{kappa}
\eea

A solution to the Boltzmann equation should be respectful with the constraints of both energy and momentum conservation, thus
\bea
 \int d\Gamma E_p\, \delta f_p 
     = \int d\Gamma {\bm p}\, \delta f_p = 0 \, , 
\label{constr}
\eea
where  $d\Gamma=d^3p/(2\pi)^3$.  Following these constraints, the authors of Ref.~\cite{Braby:2009dw} realized that with a simple linear dispersion law
the phonon contribution to the thermal conductivity would vanish. This result is consistent with the same conclusion first found out
for  superfluid helium \cite{Khala}. Thus, in this work we will compute the thermal conductivity assuming a phonon dispersion law given by 
Eq.~(\ref{NLOdisp}). The phonon
 velocitiy is therefore 
 $v_p=c_s\left(1+3\gamma p\right)$.

To evaluate  the thermal conductivity coefficient one considers the Boltzmann equation, with a collision term that takes into account the binary collisions discussed
in Sec.~\ref{phononcollisions}, and linearizes it in the deviations $\delta f_p$. After some straightforward algebra, it is possible to check that this transport coefficient can alternatively be expressed as
\bea
\kappa  = \frac{1}{12T^4}\int_{pkp'k'}
(2\pi)^4 \delta(P+K-K'-P') |{\cal M}|^2 \f f_k^0 (1+f_{k'}^0) (1+f_{p'}^0)
 \Delta_g^2 \, ,
\label{kappa2}
\eea
where $P^\mu = (E_p, {\bf p})$, and
 \bea 
 {\bm\Delta}_g 
   = g(p) {\bm p} + g(k) {\bm k} - g(k'){\bm k'} - g(p') {\bm p'} \, ,
 \label{delta_soln}
   \eea
and we have defined the shorthand
\bea
\int_q \equiv \int \frac{d^3 q}{(2\pi)^3\,2E_q}  \ .
\label{int_in_short}
\eea

The expressions given in Eqs.~(\ref{kappa}) and (\ref{kappa2}) should be equivalent, provided $g(p)$ is a solution of the Boltzmann equation.
If we use a variational approach, this equivalence  allows us to find a closed expression for $\kappa$.
We follow here the same technique and form of the variational solution  of Ref.~\cite{Braby:2009dw}.
Thus, we express the solution in terms of a basis of orthogonal polynomials
\bea
g(p) = \sum_{s=0}^{\infty} b_s B_s(p^2)\, , 
\label{g_exp}
\eea
where $B_s(p^2)$ is a polynomial in $p^2$ of order $s$. The coefficient of the lowest power is set to $1$, $B_0=1$.
 The polynomials $B_s(p^2)$ are orthogonal with each other,
\bea
\int d\Gamma \f(1+\f) p^2 B_s(p^2) B_t(p^2) \equiv A_s \delta_{st} \, .
\label{Bnorm}
\eea
With the help of $B_0=1$ all higher polynomials can be solved.

Further, demanding that the two expressions of the thermal conductivity be equivalent allows one to determine its value
as
\bea
\kappa = \left(\frac{4a_1^2}{3 T^2}\right) A_1^2 M^{-1}_{11},
\label{kappa_mod2}
\eea
where $M^{-1}_{11}$ is the (1,1) element of the inverse of the infinite matrix 
\bea
\label{M_st}
M_{st} =  \int d\Gamma_{p,k,k',p'} {\bm Q}_s \cdot {\bm Q}_t\, , 
\hspace{0.5cm}
{\bm Q}_s=  B_s(p^2) {\bm p} + B_s(k^2) {\bm k} 
              - B_s(k'^2) {\bm k'} - B_s(p'^2) {\bm p'}\ ,
             \label{kappa3}
\eea
with
\bea
\int d\Gamma_{p,k,k',p'} \equiv \int_{pkp'k'}
(2\pi)^4 \delta(P+K-K'-P') |{\cal M}|^2 \f f_k^0 (1+f_{k'}^0) (1+f_{p'}^0)  \ ,
\eea 
and for the superfluid phonons with a non-linear dispersion law one has 
\bea
a_1= \frac{4 c_s^4}{15 \Delta^2} \ ,  \qquad  A_1 = \frac{256 \pi^6}{245 c_s^9} T^9 \ .
\eea

While the matrix $M$ is a matrix of infinite dimensions, the variational treatment is performed by restricting the dimension
of the matrix to be  of order $N \times N$. Then one can prove that
\bea
\kappa \geq \left(\frac{4a_1^2}{3 T^2}\right) A_1^2 M^{-1}_{11},
\label{kappa_var}
\eea
where now $M$ is a $N \times N$ dimensional matrix. The number $N$ is treated then as a variational parameter in our numerical study.

\subsection{Temperature dependence of the thermal conductivity}
\label{ScalingSec}

Once the scattering matrix ${\cal M}$  is known, it is possible to determine the scaling behavior of $\kappa$ with the temperature and the gap, depending
whether small or large angle collisions give the leading contribution (for a similar discussion for the bulk viscosity please see Sec. 4 of
Ref.~\cite{Manuel:2013bwa}). From Eq.~(\ref{kappa_var}) it is possible to infer that
$\kappa \propto \frac{T^{16}}{\Delta^4} M^{-1}_{11}$. Assuming $N=1$ (for higher values of $N$ a similar reasoning gives the same results)
the scaling behavior of $M_{11}$ is obtained by defining dimensionless variables $x_p = \frac{c_s p}{T}$, etc. Using these dimensionless variables, also to
express the scattering matrix,
the $T$ dependence of the integrals appearing in Eq.~(\ref{M_st}) factorizes. In particular, for large angle collisions it is possible to check that
$|{\cal M}|^2 \propto T^8$, and thus
\be
\kappa \propto \frac{1}{T^2} \frac{1}{\Delta^4} \qquad  {\rm for \ large \ angle  \ collisions} . 
\ee
For small angle collisions the scaling behavior of the scattering matrix is different, as then the phonon propagator in the amplitudes of binary collisions mediated by phonon
exchange has to be considered in the collinear limit, and it is dominated by the non-linear piece in the phonon dispersion law. In this collinear limit the phonon propagator then scales as 
$1/( \gamma T^4 )\propto 1/(T^4/\Delta^2)$ instead of $1/T^2$ as in the large angle case (see Ref.~\cite{Manuel:2013bwa} for a more detailed explanation).  The change of scaling of the photon propagator then leads to
\be
\kappa \propto T^2 \frac{1}{\Delta^8} \qquad  {\rm for \ small \ angle \ collisions} .
\ee

It is still possible to consider the scaling behavior for $|{\cal M}|^2$, where ${\cal M}$ and ${\cal M}^*$ corresponds to a  small and large angle collision, respectively, or vice versa. Then the scaling arguments tells us
that 
\be
\label{combinedscaling}
\kappa \propto  \frac{1}{\Delta^6} \qquad  {\rm for \ combined \ large-small \ angle \ collisions} .
\ee
We note here that in the color-flavor-locked superfluid an explicit computation of the phonon contribution to the thermal conductivity gives the $T$-independent behavior mentioned above \cite{Braby:2009dw},
while for the superfluid  Fermi gas in the unitarity limit the phonon contribution to the thermal conductivity is dominated by small angle collisions \cite{Braby:2010ec}, although in that case $\gamma > 0$. In our case, a numerical study is mandatory in order to obtain both the $T$ scaling behavior and numerical value of the thermal conductivity for superfluid neutron matter. We present the results of such an analysis in the following
Subsection.


\subsection{Numerical values for the phonon contribution to the thermal conductivity in superfluid neutron matter}
\label{numericalsec}

\begin{figure}
\begin{center}
\includegraphics[width=0.5\textwidth]{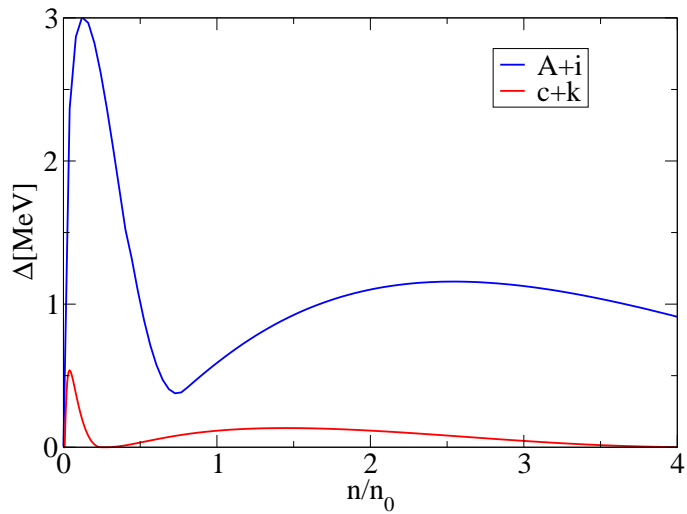}
 \caption{(Color online) The $^1S_0$ and averaged $^3P_2$ neutron gaps as a function of the nucleon particle density in units of saturation density, $n_0=0.16 \ {\rm fm^{-3}}$. 
 We use two very different gap models as a function of the density to illustrate the gap dependence of our results. In the first scheme we consider the $^1S_0 (A)$$+$$^3P_2 (i)$ model, where  the $^1S_0$ neutron gap  is calculated in the BCS approach using different bare nucleon--nucleon interactions that converge towards a maximum neutron gap of about 3 MeV at $p_F \approx 0.85 {\rm fm}^{-1}$ (parametrization $A$ of Table I in Ref.~\cite{Andersson:2004aa}) while for $^3P_2$ we have taken the parametrization $i$ (strong neutron superfluidity in the core). 
 The second model is the $^1S_0(c)$$+$$^3P_2(k)$, where for $^1S_0$ channel the calculation of the gap goes beyond the BCS theory and in the $^3P_2$ channel
we have taken into account a parametrization assuming weak neutron superfluidity in the core with maximum values for the gap of the order of 0.1 MeV \cite{Andersson:2004aa}. }
 \label{fig:gap}
\end{center}
\end{figure}

In order to calculate the thermal conductivity due to phonons in superfluid $\beta$-stable neutron matter two essential ingredients are needed, the EoS of $\beta$-equilibrated neutron star matter and the value of the gap. For the first, as already mentioned in Sec.~\ref{Intro}, we use the APR EoS for $\beta$-stable neutron matter.  The other crucial ingredient is the value of the gap of superfluid neutron matter for densities inside the core of a neutron star. 

The gap strongly depends on the nucleon-nucleon interaction model as well as on the many-body method used to compute it. Thus, there is not a consensus on its value and density dependence.  In this work we use the $^1S_0$ and averaged $^3P_2$ neutron gaps shown in Fig.~\ref{fig:gap}. We consider two very different gap models as a function of the density in order to illustrate the gap dependence of our results.  

Our first model takes into account the maximum values that the gap for superfluid neutron matter could have in the relevant $^1S_0$ and $^3P_2$ channels. The gap model, named hereafter $^1S_0(A)$$+$$^3P_2(i)$,  consists of the $^1S_0$ neutron gap that results from the BCS approach using different bare nucleon--nucleon interactions that converge towards a maximum gap of about 3 MeV at $p_F \approx 0.85 {\rm fm}^{-1}$ (parametrization $A$ of Table I in Ref.~\cite{Andersson:2004aa}). The anisotropic $^3P_2$ neutron gap is more challenging and not fully understood as one must extend BCS theory and calculate several coupled equations while including relativistic effects, since the gap extends for densities inside the core. We have taken the parametrization $i$ (strong neutron superfluidity in the core) of Table I in Ref.~\cite{Andersson:2004aa} for the $^3P_2$ neutron angular averaged value, which presents a maximum value for the gap of approximately 1 MeV. 

The second model considered is the $^1S_0(c)$$+$$^3P_2(k)$ scheme \cite{Andersson:2004aa}. In this case, for $^1S_0$ channel, the calculation of the gap goes beyond the BCS theory by adding corrections to the bare nucleon-nucleon potential such as introducing an effective neutron mass modified by the presence of the small proton fraction. For the $^3P_2$ channel we consider a parametrization assuming weak neutron superfluidity in the core with maximum values for the gap of the order of 0.1 MeV. The reason for choosing this last model lies on the fact that values for the gap of 0.1 MeV were claimed to be inferred from the fast cooling of CasA \cite{Heinke:2010cr}, although later analysis have questioned  these results \cite{Posselt:2013xva,Elshamouty:2013nfa}.

\begin{figure}
\begin{center}
\includegraphics[width=0.6\textwidth]{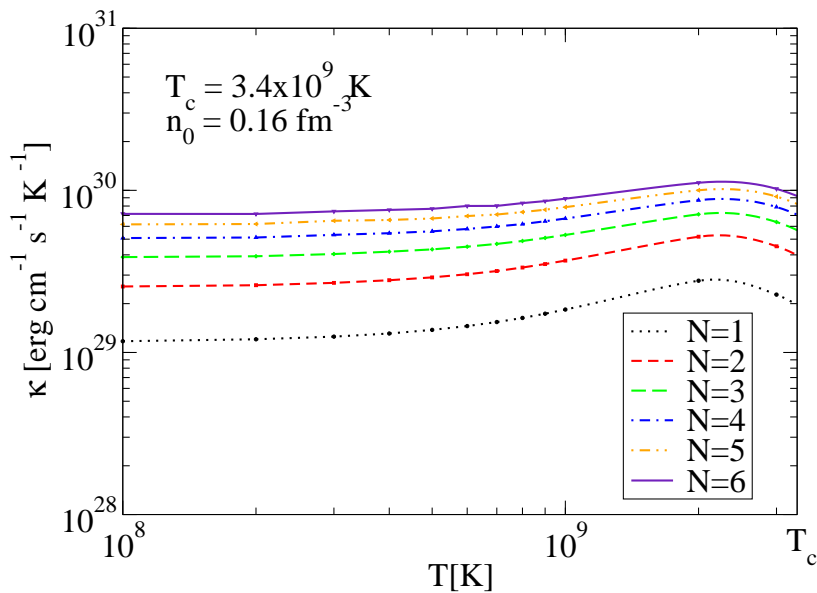}
 \caption{(Color online) Variational calculation for the thermal conductivity due to superfluid phonons up to order $N=6$ for nuclear matter saturation density, $n_0$, as a function of temperature. The end temperature is the critical temperature which amounts for $T_c=0.57 \Delta(n_0)=3.4 \times 10^9$ K. The calculation is performed  for the $^1S_0(A)$$+$$^3P_2(i)$ model for the gap.}
 \label{fig:var}
\end{center}
\end{figure}

With all these ingredients, we can proceed to calculate the thermal conductivity due to superfluid phonons using the variational method described in Sec.~\ref{sec:variational}. The evaluation of the thermally weighted scattering matrix for phonons in Eq.~(\ref{kappa3}) is performed numerically using the $Vegas$ Monte Carlo algorithm \cite{lepage1}. In Fig.~\ref{fig:var} we show the results up to order $N=6$ for nuclear matter saturation density, $n_0=0.16 {\rm fm^{-3}}$, as a function of temperature. The final value of the number $N$ is determined by imposing that the deviation with respect to the previous order should be $\lesssim 10 \%$. The end temperature is the critical temperature which amounts for $T_c=0.57 \Delta(n_0)=3.4 \times 10^9$ K, with $\Delta(n)$ given in Fig.~\ref{fig:gap} for the $^1S_0(A)$$+$$^3P_2(i)$ model for the gap.  Similar results for thermal conductivity are obtained using the $^1S_0(c)$$+$$^3P_2(k)$ gap function.

Close to $T_c$ we would expect that higher order corrections in the energy and momentum expansion should be taken into account both in the phonon dispersion law and self-interactions.
For T$ \lesssim 10^9$ K, below $T_c$, the thermal conductivity is almost independent of the temperature, with subleading temperature corrections of order $T$ and $T^2$, respectively.  In the particular case of Fig.~\ref{fig:var} and $N=6$, we find that
a fit to our numerical results is given by $\kappa \sim ( 7.02 \times 10^{29} + 9.28 \times 10^{19} \ T + 9.08 \times 10^{10} \ T^2) \ {\rm erg \ cm^{-1} \ s^{-1}  \ K^{-1}}$, with $T$ given in Kelvin. In fact, the comparison between our numerical results for the thermal conductivity for both gaps and the previous scaling arguments for the thermal conductivity with the temperature show that the dominant processes to the phonon contribution to the thermal conductivity at low T are those corresponding to the combined small and large angle collisions,  as discussed in Eq.~(\ref{combinedscaling}),
leading to an almost independent behaviour of the thermal conductivity with temperature. Then, the thermal conductivity scales as $\kappa \propto 1/\Delta^6$ at low T, the factor of proportionality depending on the density. Note that this is the same behavior found for the color-flavor-locked superfluid
\cite{Braby:2009dw}, although in that case the factor of proportionality could be expressed in terms of the quark chemical potential.

\begin{figure}
\begin{center}
\includegraphics[width=0.6\textwidth]{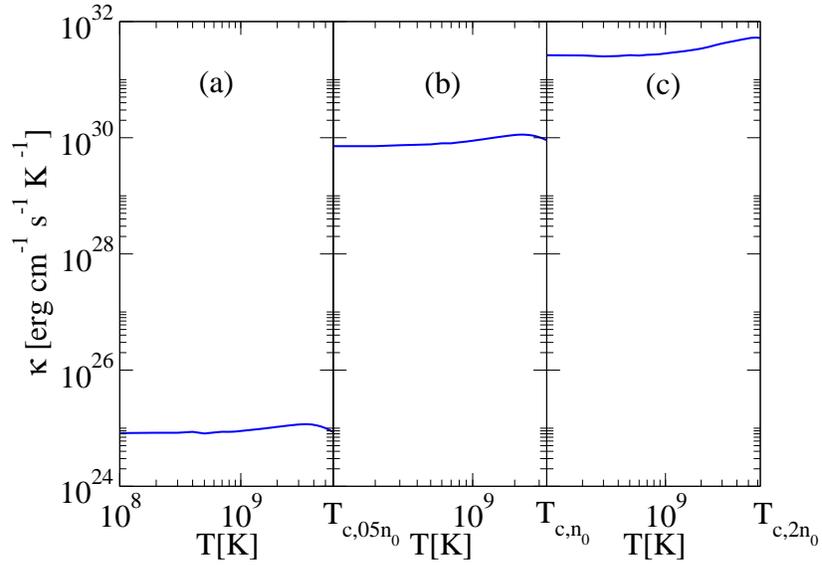}
 \caption{(Color online) Phonon contribution to the thermal conductivity using the $^1S_0(A)$$+$$^3P_2(i)$ model for the gap as a function of temperature for different particle densities in units of normal saturation density $n_0$: (a) for 0.5$n_0$ being $T_{c,0.5n_0}=5.8 \times 10^9$ K, (b) for $n_0$ being $T_{c,n_0}=3.4 \times 10^9$ K and (c) for $2n_0$ being $T_{c,2n_0}=6.3 \times 10^9$ K. The end temperature for each density is the critical temperature, $T_c=0.57 \Delta$. }
 \label{fig:difn}
\end{center}
\end{figure}

\begin{figure}
\begin{center}
\includegraphics[width=0.6\textwidth]{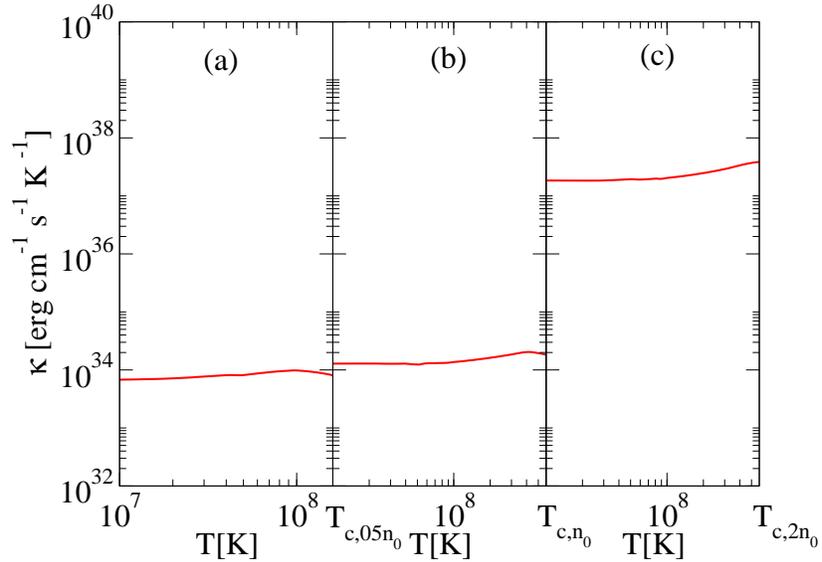}
 \caption{(Color online) The same as in Fig.~\ref{fig:difn}, but using the $^1S_0(c)$$+$$^3P_2(k)$ model for the gap. In this case, the different particle densities in units of normal saturation density $n_0$ and critical temperatures are: (a) 0.5 $n_0$ with $T_{c,0.5n_0}=1.6 \times 10^8$ K, (b) $n_0$ with $T_{c,n_0}=5.8 \times 10^8$ K and (c) $2n_0$ with $T_{c,2n_0}=5.8 \times 10^8$ K.}
 \label{fig:difn_ng}
\end{center}
\end{figure}

In Figs.~\ref{fig:difn} and \ref{fig:difn_ng} we display the contribution of phonon-phonon scattering to the thermal conductivity as a function of temperature for different particle densities in units of $n_0$ for the two models of the gap previously discussed, respectively. We start from densities of 0.5 $n_0$, as the lower density one can reach in the core of the neutron star. The end temperature for each density is the critical temperature, $T_c=0.57 \Delta$, with the gap  for different densities given in Fig.~\ref{fig:gap}.  The thermal conductivity grows with increasing density, with a non-linear dependence according to the evolution with density of the gap.

\begin{figure}
\begin{center}
\includegraphics[width=0.6\textwidth]{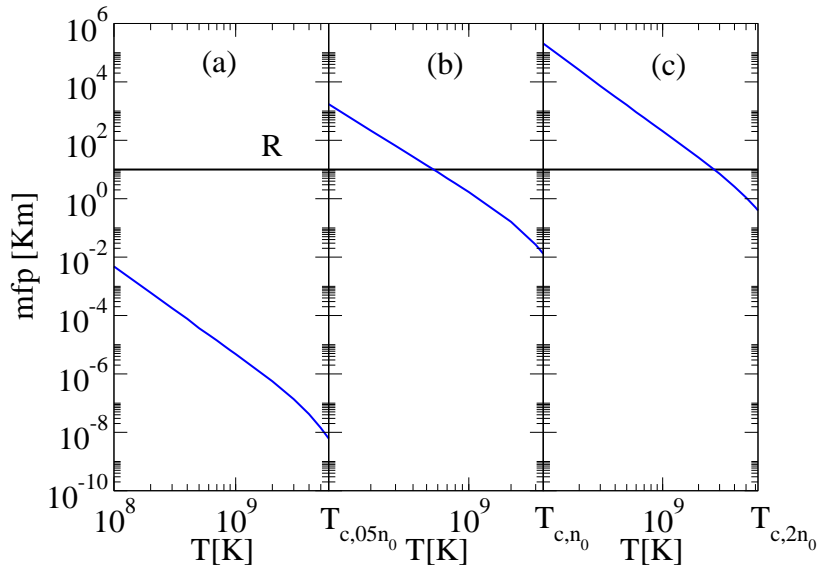}
 \caption{(Color online) Mean-free path due to the phonon thermal conductivity using the $^1S_0(A)$$+$$^3P_2(i)$ model for the gap as a function of temperature for three different densities in units of $n_0$: (a) for 0.5$n_0$ being $T_{c,0.5n_0}=5.8 \times 10^9$ K, (b) for $n_0$ being $T_{c,n_0}=3.4 \times 10^9$ K and (c) for $2n_0$ being $T_{c,2n_0}=6.3 \times 10^9$ K.  The end temperature for each density is the critical temperature, $T_c=0.57 \Delta$. The mean-free path has to be compared with the radius of the star, which we take $R=10$ Km.  }
 \label{fig:mfp}
\end{center}
\end{figure}


From our results, one can also extract the thermal conductivity mean free path of the phonons, which is not the same as the mean free path associated to shear viscosity (see Ref.~\cite{Manuel:2011ed} for a comparison). For the thermal conductivity 
the mean free path (mfp) is defined as
\bea
l= \frac{\kappa}{\frac{1}{3} c_v c_s} ,
\eea
with the heat capacity for phonons given by \cite{Khala}
\bea
c_v= \frac{2 \pi^2}{15 c_s^3} \left( T^3 + \frac{25 \gamma}{ 7} \frac{(2 \pi)^2}{c_s^2} T^5 \right) \ .
\eea
We show in Fig.~\ref{fig:mfp} the mfp of phonons in $\beta$-stable neutron star matter for $0.5 n_0$, $n_0$ and $2 n_0$ as a function of temperature using the $^1S_0(A)$$+$$^3P_2(i)$ model for the gap. 
We also indicate an estimate of the limit of 10 Km for the radius of the star. We observe that  for $n= 0.5 n_0$ the superfluid phonon mfp stays below the radius of the star, also
 for $n= n_0$ and T $\gtrsim 6\times10^8$ K  as well as for 2$n_0$ and T $\gtrsim 3\times10^9$ K. As for the $^1S_0(c)$$+$$^3P_2(k)$ gap function, we obtain that the superfluid phonon mfp for the temperatures and densities studied are too large, as the thermal conductivity is several orders
 of magnitude higher than the previous case. Thus, we only show numerical results
for our first gap. We find that $l \propto 1/T^3$. 
The temperature dependence of the mfp results from the almost temperature independent behaviour of the thermal conductivity.

A relevant discussion is to know the temperature regime where our  hydrodynamical results are valid.
As mentioned previously,
 close to $T_c$ we would expect that higher order corrections in the energy and momentum expansion should be taken into account both in the phonon dispersion law and self-interactions, but these have not been
derived yet. Further, the density of superfluid phonons becomes very dilute at very low $T$, and then it might be difficult to maintain a hydrodynamical description of their behavior, if the phonons do not collide with other light particles, such as the electrons.
Otherwise, the phonons would behave in the low $T$ regime ballistically. In such a case the thermal conductivity due to phonons would be then dominated by the collisions of the phonons with the boundary.
 The behavior of the thermal conductivity in this regime can be qualitatively estimated as
\bea
\kappa = \frac{1}{3} c_v c_s R ,
\eea
with $R$ being the radius of the superfluid core of the star. In this case $\kappa \propto T^3$, and it would clearly drop at sufficiently low $T$. This ballistic behavior of the phonons of superfluid $^4$He has been experimentally tested \cite{helio}.  However, the process of heat transport in a ballistic regime follows a different law than Eq.~(\ref{defi}) (see Refs.~\cite{chen,alvarez,christov}). In such a case, a more detailed analysis of the Boltzmann equation obeyed by the phonons should be carried out. Prior to this, one should analyze whether electron-superfluid phonon interactions could be more important than the phonon collisions with the boundary and whether these may give a relevant contribution to the thermal conductivity. This will be the subject of future studies.


\section{Conclusions}
\label{conclusions}

We have carried out a numerical computation of the phonon contribution to the thermal conductivity  in superfluid $\beta$-stable neutron matter. We have obtained the phonon scattering rates using effective field theory techniques in terms of the EoS of the system. For that purpose, we have used a causal parametrization of a widely used benchmark, the APR EoS.  Moreover, since the thermal conductivity vanishes using a linear phonon dispersion law, we have calculated the first correction beyond linear order, which depends on the gap of neutron matter. Two different models for the gap in the strong and weak superfluidity regimes have been considered. With all these ingredients, we have solved the Boltzmann equation using a variational approach and  have checked that, as it occurs for the color-flavor locked superfluid, it can be well approximated  for $T \ll T_c$
by an almost $T$ independent result, where  $\kappa \propto 1/\Delta^6$, the factor of proportionality depending on the density and  EoS of the superfluid. Our analysis indicates that this behavior arises by the combined effect
of large and small angle phonon binary collisions.

Our results should be compared to the thermal conductivity of electrons and muons mediated by electromagnetic interactions in neutron star cores. 
In Ref.~\cite{Shternin:2007ee} the calculations of the electron contribution to the thermal conductivity has been revised, considering 
 the effect of the Landau damping of electromagnetic interactions owing to the exchange of transverse plasmons as well as the presence of proton superconductivity  that had been previously neglected \cite{Flowers,Baiko:2001cj}.
  According to Fig.~5 of Ref.~\cite{Shternin:2007ee}, for a mass density of $\rho=4 \times 10^{14} {\rm g \ cm}^{-3}$, the thermal conductivity of electrons and muons, $\kappa_{e \mu}$, has values of $\kappa_{e \mu} \sim 10^{23-24} \  {\rm erg \ cm^{-1} \ s^{-1}  \ K^{-1}}$ for $T \sim 10^8$ K depending on the strength of the proton superconductivity, while $\kappa_{e \mu}$ is reduced by few orders of magnitude at $T \sim 10^9$ K following  the asymptotic law, $\kappa_{e \mu} \propto T^{-1}$.  For lower mass densities and in the limit of strong superfluidity, we could easily estimate the thermal conductivity using Eq.~(93) of Ref.~\cite{Shternin:2007ee}. In this case we find that for a mass density of $\rho \sim 1.4 \times 10^{14} {\rm g \ cm}^{-3}$ (that corresponds to the lowest density inside the core of $n \sim 0.5 n_0$) the thermal conductivity of electrons and muons is reduced by a factor three at most. Thus, we find that thermal conductivity in the neutron star core is dominated by phonon-phonon collisions when phonons 
  are in  a pure  hydrodynamical regime, since $10^{25} \lesssim \kappa_{ph} \lesssim 10^{32}  \  {\rm erg \ cm^{-1} \ s^{-1}  \ K^{-1}}$ from 0.5$n_0$ to 2$n_0$ for temperatures up to $T_c$.  
  
 Our results also indicate that if the  contribution of electrons-muons and phonons to $\kappa$ get comparable, electron-phonon collisions could play an important role in the determination of $\kappa$. It is possible to study those interactions using also the effective field theory techniques \cite{Manuel:2012rd}. If the electron-phonon collisions are relevant, one would then be  forced to perform a study of the thermal conductivity in an electron-phonon coupled system. Simple estimates have been done in Ref.~\cite{Bedaque:2013fja}. This topic should be carefully analyzed in future studies.


\appendix

\section{Polarization tensor}

\label{App:Polarization}

We show here some intermediate steps and details for the computation phonon dispersion law.
The computation of $\Pi_{\mu \nu}$ is similar to the computation of the one-loop polarization tensor associated
to a $U(1)$ gauge field (see also Ref.~\cite{Schafer:2006yf}). Starting from Eq.~(\ref{pimunu}), we perform the 
 trace over the indices in the Nambu-Gorkov space. Then one can express the different (symmetric) components
of the polarization tensor as follows
 \ba
\label{pi00_ng}
\Pi_{00}(q_0, {\bf q} ) & = &-2i 
  \int\!\frac{d^4p}{(2\pi)^4} 
   \left\{\frac{p_0k_0+\xi_p\xi_k-\Delta^2}{\left(p_0^2-\xi_p^2-\Delta^2\right)
   \left(k_0^2-\xi_k^2-\Delta^2\right)}\right\},
 \\
 \label{pi_ij_ng}
\Pi_{ij}(q_0,  {\bf q} ) & = &-2i 
  \int\!\frac{dp_0}{(2\pi)}\frac{d^3p}{(2\pi)^3} 
   \left\{v_iv_j\frac{p_0k_0+\xi_p\xi_k+\Delta^2}{\left(p_0^2-\xi_p^2-\Delta^2\right)
   \left(k_0^2-\xi_k^2-\Delta^2\right)}\right\},
\\
\label{pi_0i_ng}
\Pi_{0i}(q_0,  {\bf q} ) & = &-2i
  \int\!\frac{dp_0}{(2\pi)}\frac{d^3p}{(2\pi)^3} v_i
   \left\{\frac{p_0\xi_k+k_0\xi_p}{\left(p_0^2-\xi_p^2-\Delta^2\right)\left(k_0^2-\xi_k^2-\Delta^2\right)}\right\} ,
\ea  
where $k_0 = p_0-q_0$ and $\xi_k =\xi_{(p-q)}$, and
we have considered that the velocities are  ${\bf v}_{k}={\bf v}_{(p-q)}={\bf v}$ .

The $p_0$ integral can be performed by contour integration, assuming
retarded boundary conditions $p_0 \rightarrow p_0 + i 0^+ {\rm sign}(p_0)$. The remaining angular integrals can be easily performed if we only compute
$q^\mu q^\nu \Pi_{\mu \nu}(q_0,q)$. The final integral in $p$ can then easily performed if we assume that  $\mu/\Delta\rightarrow \infty$, which is a good approximation 
given the values of the neutron gaps and chemical potential. Then one finds
\ba
\Pi_{00}(q_0, q) &= &\frac{2p_F  m^*}{\pi^2}\left(1+\frac{q_0^2}{6\Delta^2} + {\cal O} \left( \frac{q_0^4}{\Delta ^4},  \frac{q^4 v_F^4}{\Delta ^4} \right)  \right) \ ,
\label{Pi_00}
\\
 q^iq^j\Pi_{ij}(q_0, q) &= &\frac{p_F m^*q^2v_F^2}{30\pi^2}\left(\frac{q^2v_F^2}{\Delta ^2}+   {\cal O} \left( \frac{q_0^4}{\Delta ^4},  \frac{q^4 v_F^4}{\Delta ^4} \right)
 \right) \ ,
 \label{pi_ij}
 \\
q^i\Pi_{0i}(q_0, q) &= &\frac{p_F m^* q v_F}{2\pi^2}\left(\frac{1}{9} \frac{q_0qv_F}{\Delta^2}+ {\cal O} \left( \frac{q_0^4}{\Delta ^4},  \frac{q^4 v_F^4}{\Delta ^4} \right)  \right) \ ,
 \label{pi_0i}
 \ea
which are the values needed to compute the phonon dispersion law at the order of accuracy required in this manuscript.

 \acknowledgments{ We thank M. Alford and A. Schmitt for useful discussions on the topic of this manuscript.
 This research was supported by Ministerio de Econom\'{\i}a y Competitividad under contracts FPA2010-16963 and FPA2013-43425-P, as well as ÒNewCompStarÓ (COST Action MP1304).
  LT acknowledges support from the Ramon y Cajal Research Programme from Ministerio de Econom\'{\i}a y Competitividad and from FP7-PEOPLE-2011-CIG under Contract No. PCIG09-GA-2011-291679.}

\end{document}